# IEEE Copyright Notice

# IMPROVING PROSTATE WHOLE GLAND SEGMENTATION IN T2-WEIGHTED MRI WITH SYNTHETICALLY GENERATED DATA

*Alvaro Fernandez-Quilez*[1,2,†]    *Steinar Valle Larsen*[2,3,†]

Morten Goodwin [4]    Thor Ole Gulsrud [1]    Svein Reidar Kjosavik [5]    Ketil Oppedal[2,3,6]

[1]Department of Quality and Health Technology, University of Stavanger, Norway.
[2]Stavanger Medical Imaging Laboratory (SMIL), Stavanger University Hospital, Norway.
[3] Department of Electrical Engineering and Computer Science, University of Stavanger, Norway.
[4] Department of ICT, University of Agder, Grimstad, Norway.
[5] General Practice and Care Coordination Research Group, Stavanger University Hospital, Norway.
[6] Centre for Age-Related Medicine, Stavanger University Hospital, Norway.

## ABSTRACT

Whole gland (WG) segmentation of the prostate plays a crucial role in detection, staging and treatment planning of prostate cancer (PCa). Despite promise shown by deep learning (DL) methods, they rely on the availability of a considerable amount of annotated data. Augmentation techniques such as translation and rotation of images present an alternative to increase data availability. Nevertheless, the amount of information provided by the transformed data is limited due to the correlation between the generated data and the original. Based on the recent success of generative adversarial networks (GAN) in producing synthetic images for other domains as well as in the medical domain, we present a pipeline to generate WG segmentation masks and synthesize T2-weighted MRI of the prostate based on a publicly available multi-center dataset. Following, we use the generated data as a form of data augmentation. Results show an improvement in the quality of the WG segmentation when compared to standard augmentation techniques.

## 1. INTRODUCTION

Prostate cancer (PCa) is the second most common diagnosed cancer [1], with an estimated incidence of 1.3 million new cases among men worldwide in 2018 [2, 3].

Thanks to recent advances in image acquisition and interpretation, MRI has proven to be a valuable tool for PCa detection, staging, treatment planning and intervention [4]. Segmentation of the prostate from MRI plays a crucial role. For instance, radiotherapy planning, MRI-transrectal ultrasound fusion guided biopsy or radiation dose planning in brachytherapy are highly dependent on an accurate delineation of the prostate in imaging data [5]. Current practices include manual contouring by a specialist in a slice-by-slice basis, which is a time and labor intensive task as well as susceptible to intraobserver and interobserver variability.

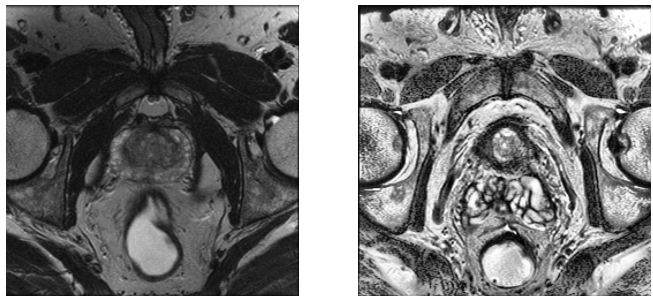

**Fig. 1**: Example of an original slice (left side) and preprocessed slice (right side).

In recent years, convolutional neural networks (CNNs) have shown promise in segmentation, where U-net was a major breakthrough [6]. Nevertheless, without further refinement, the training of CNN-based methods require a considerable amount of data [7]. Moreover, imbalanced data or data with low variability might lead to sub-optimal results [8]. Augmentation techniques (e.g. translation and rotation) have proven useful but they also produce highly correlated data, limiting the amount of information provided to the algorithm in the training phase.

Generative Adversarial Networks (GAN) [9] have gained a considerable amount of attention in the DL community. Several variations of these generative models have been developed, such as the deep convolutional GAN (DCGAN) [10] or pix2pix [11] which are able to generate realistic images after learning the distribution of the original dataset and have been used to generate T1-weighted brain MRI [12]. The pix2pix architecture has been used to translate brain masks to images [13]. We propose a pipeline to generate paired prostate masks and T2-weighted MRI of the prostate for data augmentation. Our contributions in this work are:
1. We propose a DCGAN-based architecture to generate whole gland prostate masks from T2-weighted MRI.
2. We propose a pix2pix-based architecture to translate the synthetic WG prostate MRI masks into T2-weighted prostate MRI and to obtain paired training samples.
3. We provide a comprehensive comparison of different data augmentation techniques and their effect on the WG segmentation of the prostate as well as the effect of adding synthetic data to those standard augmentation techniques.

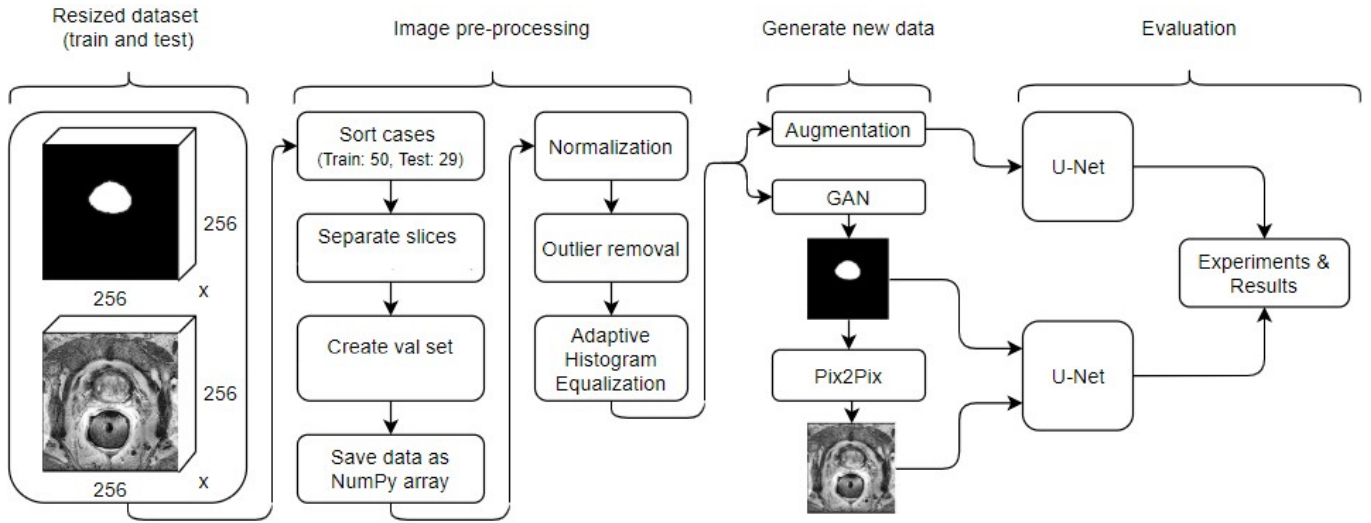

**Fig. 2**: Technical approach to the project.

## 2. METHODS

In this work, we propose a semi-automatic pipeline able to generate synthetic pairs of T2-weighted prostate MRI and their respective WG mask. Figure 2 presents an overview of the steps followed in the work. First, we provide a description of the dataset, the pre-processing steps and the architectures training process.

### 2.1. Dataset

We use the PROMISE12 data set [14], containing T2-weighted axial MRI of 50 patients for training and 30 for testing. Ground truths of the WG of the prostate annotated by the experts are only available in the training set whilst the testing ones can only be accessed when submitting the results[1].

[1]For up-to-date information refer to https://promise12.grand-challenge.org/

### 2.1.1. Preprocessing and data splitting

We perform four different steps. First, MRI are re-sampled by linear interpolation to 256x256, which is the lowest resolution present in the data set. Following, the intensity is then normalized to an interval of [0,1]. Outlier removal is performed by forcing the pixel intensity values of the image between the 1st and 99th percentiles. Finally, a contrast limited adaptative histogram equalization (CLAHE) is applied to improve local contrast and enhance the edge definition [15]. Figure 1 shows an example of the result obtained after applying the preprocessing steps. In order to evaluate the methods, we split the original dataset by patients following a 60%/20%/20% split for the training, validation and testing set, respectively. Eventually, an independent assessment is done using the PROMISE12 test set.

### 2.2. Segmentation architecture

Our segmentation architecture is based on the original U-net architecture [6].

### 2.2.1. Training of the network

Dice coefficient (DSC) [16] is used to evaluate the model performance during training. The architecture is trained for 200 epochs with a batch size of 32 on a 16gb NVIDIA Tesla P100, based on the validation set. Adam optimizer [17] is used with a learning rate scheduler. The learning rate started with a value of 1e-3 and reduced by a factor of 10 if the loss did not decrease during 10 epochs.

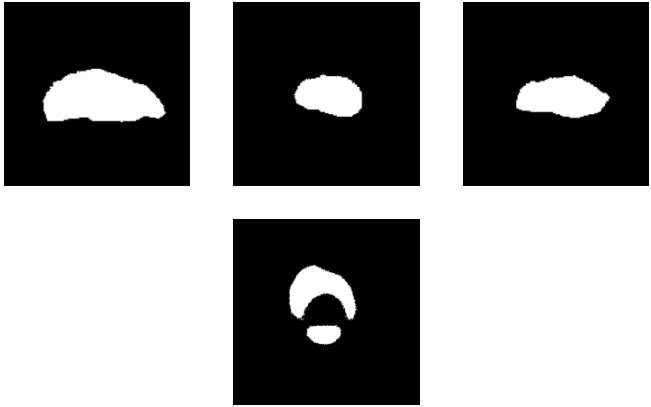

**Fig. 3**: Realistically-looking masks generated by DCGAN (top row) and a mask deemed as unrealistic (bottom).

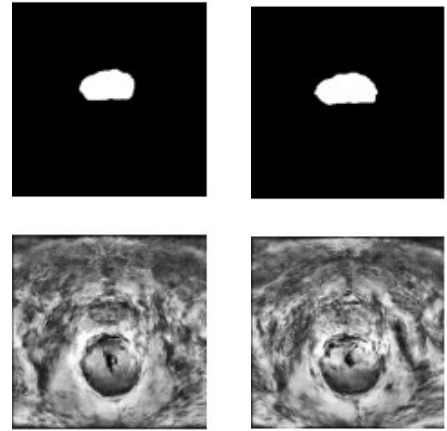

**Fig. 4**: Synthesized T2 weighted MRI (bottom) from GAN-generated WG masks (top row).

### 2.3. Generation of synthetic WG masks

We adopt the DCGAN architecture [10] to generate synthetic WG prostate masks. The DCGAN consists of two main components, a generator ($G$) and a discriminator ($D$), where both G and $D$ are CNNs. The generator $G$ gets inputs samples $z$ from a distribution which has a normal prior imposed to it $pz \sim \mathcal{N}(\mu, \sigma^2)$. The role of $G$ is to map such samples to the original data space while inducing a specific distribution *pdata* and thus synthesizing samples that follow such distribution $\hat{x} = G(z)$). On the other hand, the role of $D$ is to discriminate between real data samples x and generated ones $\hat{x}$.

#### 2.3.1. Training of the network

The generator network takes a vector $z$ of 100 random numbers drawn from a normal distribution $p_z \sim \mathcal{N}(\mu, \sigma^2)$ as inputs and outputs 256x256 WG segmentation masks. The generator architecture follows the one proposed in [10]. Nevertheless, after an extensive testing, we added two transposed convolutional layers. In addition, the original paper makes use of a rectified linear unit (ReLU) activation for all the generator layers except for the output whereas this work makes use of the LeakyReLU [18]. The discriminator implementation follows, again, the original one [10] with the addition of two extra convolutional layers. The architecture is trained for 1500 epochs on a 16gb NVIDIA Tesla P100, based on the validation set. Each epoch took approximately 100 seconds to finish. The batch size used for the training was 32. Adam optimizer is used with default parameters and a learning rate of 0.0002.

#### 2.3.2. Mask selection

The selection criteria for the synthetic masks is based on the visual appearance of the image and done in a manual way. For instance, some of the synthetic masks might contain disconnected prostate glands. Figure 3 shows an example of realistic WG prostate masks as well as an unrealistic one.

### 2.4. Mask-to-image translation: T2-weighted MRI

We base our architecture on [11]. In the particular case of translation, the generator ($G$) aims to map a source domain image $x_s \sim p_{xs}$ into the corresponding target image $x_t \sim p_{xt}$ via the mapping function $G(x_s, x_t)$. In this case, the discriminator tries to discriminate between the source image and its corresponding ground truth by classifying them as real while classifying the input and the transformation as fake.

#### 2.4.1. Training of the network

The input of the architecture is a 256x256 segmentation mask while the output was a synthesized T2-weighted MRI from the input mask. Figure 4 shows two examples of synthesized T2-weighted prostate MRI. We train the network for 200 epochs with a batch size of 1 on an Nvidia Tesla P100 16gb, based on the validation set. Each epoch took approximately 300 seconds to finish. The optimizer is Adam with a learning rate of 0.0002.

### 3. RESULTS

We evaluate the segmentation results following the metrics used in the PROMISE12 challenge as well as additional ones: DSC, mean volumetric DSC (VDSC), mean surface distance (MSD) and mean hausdorff distance (HD). The quality of the synthetic data was manually evaluated in a visual way in an intermediate step. The segmentation metrics are both representative of slice-based metrics as well as volume-based. More details on the metrics can be found in [14, 19]. Standard augmentation included rotation ($\pm 10$ degrees), shifting (10% total height and width), flipping and zooming ([1, 1.2] range). All results are based on U-net and the effect of the segmentation architecture was considered to be out of the scope of the paper. Generated data results are based on 10000 synthetic T2-weighted images and their corresponding masks, which is approximately 8 times the amount of original data. Quantitative results on the usage of synthetic data as well as standard augmentation techniques can be found in table 1 while table 2 depicts the effect of the combination of the standard augmentation techniques and synthetically generated data.

**Table 1**: Standard augmentation techniques and synthetic data effect on WG segmentation results.

| Transformation | DSC (%) | MSD | HD | VDSC(%) |
|---|---|---|---|---|
| Original | 67.84 | 3.61 | 8.86 | 54.30 |
| Vertical flip | 66.93 | 18.52 | 19.68 | 48.72 |
| Horizontal flip | 69.98 | 15.41 | 13.47 | 50.86 |
| Rotation | 73.07 | 3.59 | 8.53 | 59.78 |
| Shift | 71.33 | 12.24 | 9.44 | 56.16 |
| Zoom | 70.69 | 7.31 | **7.74** | 55.21 |
| All | 67.30 | 10.14 | 12.36 | 51.23 |
| Synthetic data | **73.77** | **1.16** | 8.10 | **69.36** |

**Table 2**: Combination of standard augmentation techniques with synthetic data.

| Transformation | DSC(%) | MSD | HD | VDSC(%) |
|---|---|---|---|---|
| Original | 67.84 | 3.61 | 8.86 | 54.30 |
| Vertical flip | 67.50 | **0.92** | 12.80 | 68.27 |
| Horizontal flip | 72.84 | 1.40 | 7.02 | 69.79 |
| Rotation | 68.01 | 1.93 | 9.93 | 68.06 |
| Shift | 73.37 | 1.18 | 8.66 | **73.32** |
| Zoom | **73.90** | 1.56 | **6.94** | 70.90 |
| All | 69.81 | 1.60 | 7.99 | 66.83 |
| Synthetic data | 73.77 | 1.16 | 8.10 | 69.36 |

A mean DSC of 73.77% was obtained for the WG of the prostate when adding synthetically generated data, both masks and the T2-weighted synthesized MRI. On the other hand, a DSC of 67.84% was obtained with a vanilla U-net without any augmentation. When comparing HD, the synthetically augmented dataset also improved the vanilla U-net results (8.86 mm) by more than 8%. In addition, the MSD has also shown a considerable improvement when using a synthetically augmented dataset. Finally, the volumetric DSC also increased by more than 15% when using synthetic data. When comparing standard augmentation with the synthetic one, the latest surpasses by a considerable margin all the other techniques when it comes to MSD and VDSC as well as by a small margin the DSC. Amongst the standard augmentation techniques rotation obtained the best results.

Using synthetic data in combination with standard augmentation techniques yield to a larger improvement over the previous results. The combination resulted in an improvement of the metrics for all the standard augmentation techniques with the exception of rotation. In particular, zoom showed the best results amongst all of them with improvements with respect to the standard augmentation as well as the baseline.

Further experiments where performed to explore the effect of the synthetic sample size on the DSC value. Different % of synthetic data with respect to the size of the original sample were tested and we found out that the increase in the results was consistent with the amount of synthetic data used to augment the original set. In particular, we observed that every time the amount of synthetic data was doubled the DSC increased around 2%, reaching the peak with the largest amount of data tested (10000 cases).

## 4. CONCLUSIONS

The objective of this work is to provide a pipeline able to provide an alternative to the standard augmentation techniques by making use of GAN-based architectures. We propose a GAN-based framework to generate prostate WG segmentation masks and synthesize T2-weighted MRI from them. We evaluate our method against the standard augmentation techniques while providing a comprehensive comparison of the effect of them in the prostate WG segmentation in T2 weighted MRI. Results have shown that our method was able to obtain better results when used as a standalone technique than the standard augmentation techniques whilst also improving the results of the standard augmentation techniques when used in combination with them.

To the best of our knowledge, this is the first approach that makes use of generated synthetic T2 weighted prostate MRI and their generated paired WG masks to improve WG segmentation. Whilst our method showed promise improving WG prostate segmentation results, the framework would benefit from inclusion of an automatic way to asses the quality of the generated images and select those deemed as more realistic. Furthermore, refinement of the synthesis architecture (pix2pix) could be explored in order to have more realistic details and less blurriness around the gland boundary.

## 5. COMPLIANCE WITH ETHICAL STANDARDS

## 6. ACKNOWLEDGMENTS

This work has been funded by the University of Stavanger. The authors have no relevant financial or non-financial interests to disclose.